# Multi-focus Image Fusion for Visual Sensor Networks

Milad Abdollahzadeh, Touba Malekzadeh, Hadi Seyedarabi
Faculty of Electrical and Computer Engineering
University of Tabriz
Tabriz, Iran
Email: {milad.abdollahzadeh, seyedarabi}@tabrizu.ac.ir, t.malekzadeh94@ms.tabrizu.ac.ir

*Abstract*— **Image fusion in visual sensor networks (VSNs) aims to combine information from multiple images of the same scene in order to transform a single image with more information. Image fusion methods based on discrete cosine transform (DCT) are less complex and time saving in DCT based standards of image and video which makes them more suitable for VSN applications. In this paper an efficient algorithm to fusion of multi-focus images in DCT domain is proposed. Sum of modified laplacian (SML) of corresponding blocks of source images are used as contrast criterion and blocks with larger value of SML are absorbed to output images. The experimental results on several images show the improvement of proposed algorithm in terms of both subjective and objective quality of fused image relative to other DCT based techniques.**

## I. Introduction

Image fusion is generally defined as the process of combining multiple input images into a smaller set of images, usually a single one, which contains all the important information from the inputs [1]. The aim of image fusion is to reduce the amount of data and create new images that are more suitable for the purposes of visual perception and machine processing. Due to the limited depth of optical lenses only the objects in a particular distance are in focus. So, it is impossible to describe a complex scene with a single image precisely [2]. Multi-focus image fusion creates an image which almost all objects are in focus.

Visual sensor networks (VSNs) is the term used in the literature to refer to a system with a large number of cameras, geographically spread on monitoring points [3]. In VSNs multiple cameras obtain multiple images of a scene and a centralized fusion center combines source images to create a more informative image. Then fused image will be transmitted to an upper node [4]. In VSNs, especially when nodes are wireless, energy consumption of communication stage is more than the energy consumption of data processing. Therefore, images are compressed before transmission to the other nodes. It means that fusion center will receive compressed version of source images instead of original version.

So far, several researchers have been focused on image fusion which are performed on the images in spatial and spectral domain [5-7]. Laplacian, gradient, Discrete Wavelet Transform (DWT) [8] and Shift Invariant Discrete Wavelet Transform (SIDWT) [9] are examples of multi-scale decompositions used on image fusion. Multi-scale decompositions are complex and time consuming, so most of the image fusion approaches based on these transforms are unsuitable for VSNs due to the resource constraints.

In VSNs, when the source images are transmitted or saved with DCT based standards, DCT based image fusion algorithms will reduce complexity considerably [10]. With this in mind, recently several image fusion algorithms based on DCT transform have been proposed. Tang [11] has proposed two image fusion techniques in DCT domain, namely DCT+Average and DCT+Contrast. These methods are performed on each 8x8 block of DCT representation of source images and the block with the highest activity level has been choose. These methods are suffer from undesirable side effects like blurring or artifacts. So, the fused image has poor subjective quality. In [12] image fusion technique computes variance on DCT domain in order to reduce complexity of the algorithm. Phamila [13] extended work of [12], by choosing the blocks with higher valued AC coefficients of DCT transform. Most of the AC coefficients have small value, so they are quantized to zeros during the quantization. This leads to mistakes in selecting right JPEG coded blocks because the number of higher valued AC coefficients is an invalid criterion. However, experiments at [14] shows that variance [12] is the worst focus measures.

In this paper a general image fusion approach based on DCT transform and Sum of Modified Laplacian (SML) criteria is proposed. Blocks with higher value of SML are absorbed to the fused image. A consistency verification procedure is followed to increase quality of output image. Experimental results shows that our algorithm improves quality of fused image considerably.

The rest of this paper is organized as follows: In section 2 basic concepts of our approach are discussed and SML calculation on DCT domain is driven. Section 3 contains our proposed method for image fusion. Section 4 presents simulation results. Conclusion is made on section 5.

## II. DCT Block Analysis

### A. Discrete Cosine Transform

DCT based compression standards are most popular standards for image and video compression. Some examples include JPEG still image coding standard [15], Motion-JPEG, MPEG and the ITU H.26X video coding standards [16].

### B. SML in DCT domain as a fusion criterion

Nayar noted that in the case of the Laplacian computation the second derivatives in the x- and y-directions can have opposite signs and tend to cancel each other [17]. Therefore he proposed the Modified Laplacian (ML) as:

$$\nabla^2_M I = \left|\frac{\partial^2 I}{\partial x^2}\right| + \left|\frac{\partial^2 I}{\partial y^2}\right| \qquad (1)$$

Where $I$ and $\nabla^2_M I$ are two dimensional input signal and its ML respectively. The expression for discrete approximation of ML in order to use in digital image processing applications is:

$$\nabla^2_{ML} I(x,y) = |2I(x,y) - I(x-step,y) - I(x+step,y)| +$$
$$|2I(x,y) - I(x,y-step) - I(x,y+step)| \qquad (2)$$

Where $I(x,y)$ represent pixel value of digital image $I$ located at $xth$ row and $yth$ column and 'step' is an integer used to accommodate for possible variations in size of texture elements. After computing ML for each pixel within a $N \times N$ window of image, SML results from summation of all ML values:

$$SML = \sum_{x=0}^{N-1}\sum_{y=0}^{N-1} \nabla^2_{ML} I(x,y) \quad for \quad \nabla^2_{ML} I(x,y) \geq T \qquad (3)$$

In (3), 'T' is a discrimination threshold value. According to (5-6) SML calculation on DCT domain requires to find effect of shifting input signal on DCT coefficients. Reeves [18] investigated properties of DCT transform after applying linear operations. Based on this work, DCT coefficients of linearly transformed signal $g_L(n)$ can be obtained by DCT coefficients of original signal $g(n)$ by:

$$G_L(u) = \sum_{v=0}^{N-1} G(v) \sum_{n=0}^{N-1} \sqrt{\frac{2}{N}} \cos\left(\frac{(2n+1)u\pi}{2N}\right) r_{Lv}(n) \qquad (4)$$

Where $G_L(u)$ and $G(v)$ are DCT of $g_L(n)$ and $g(n)$ respectively. Furthermore, $r_{Lv}(n)$ can be obtained by applying linear transform on $r_v(x)$ followed by sampling at $x=n$:

$$r_v(n) = c(v)\sqrt{\frac{2}{N}}\cos(\frac{(2n+1)v\pi}{2N}) \qquad (5)$$

According to (2) each pixel of $N \times N$ window of image needs at least four shifts, two absolute value and 5 summation. So, SML calculating using (3) is a complex task and waste lots of resources. A simpler way is to use (1) to calculating ML. We need to apply second derivative operator in both $x$ and $y$ direction and sum of absolute of second derivatives results in ML. Based on separable nature of two dimensional DCT transform [18] we can apply second derivative operators on $x$ and $y$ directions separately. In order to investigate effect of second derivative on DCT coefficient, we need to obtain second derivatives of $r_u(x)$ and $r_v(y)$ and sample them at $x=n$ and $y=n$ respectively. It results in $r_{Lu}(m)$ and $r_{Lv}(n)$ kernels:

$$r_{Lu}(m) = \left(\frac{u\pi}{N}\right)^2 \sqrt{\frac{2}{N}} c(u) \cos\left(\frac{(2m+1)u\pi}{2N}\right) \qquad (6)$$

$$r_{Lv}(n) = \left(\frac{v\pi}{N}\right)^2 \sqrt{\frac{2}{N}} c(v) \cos\left(\frac{(2n+1)v\pi}{2N}\right) \qquad (7)$$

Then, effect of second derivatives of $x$ and $y$ can be investigated by columns and rows of DCT coefficients of $N \times N$ window of $I$ as follows:

$$G_x(\alpha,\beta) = \sum_{u=0}^{N-1} G(u,\beta) \times$$
$$\sum_{m=0}^{N-1} \frac{2(u\pi)^2}{N^3} c(\alpha)c(u) \cos\left(\frac{(2m+1)\alpha\pi}{2N}\right) \cos\left(\frac{(2m+1)u\pi}{2N}\right) \qquad (8)$$

$$G_y(\alpha,\beta) = \sum_{v=0}^{N-1} G(\alpha,v) \times$$
$$\sum_{n=0}^{N-1} \frac{2(u\pi)^2}{N^3} c(\beta)c(v) \cos\left(\frac{(2n+1)\beta\pi}{2N}\right) \cos\left(\frac{(2n+1)v\pi}{2N}\right) \qquad (9)$$

Where $G$ is the DCT of original image $I$ and $\alpha,\beta,u,v = 0,1,...,N-1$. Here $G_x$ and $G_y$ represent DCT of $\frac{\partial^2 I}{\partial x^2}$ and $\frac{\partial^2 I}{\partial y^2}$ respectively. Using (8) and (9) we can calculate SML of each $N \times N$ image:

$$SML = \sum_{\alpha=0}^{N-1}\sum_{\beta=0}^{N-1} (|G_x(\alpha,\beta)| + |G_y(\alpha,\beta)|) \qquad (10)$$

In conclusion SML of an $N \times N$ block of pixels can be computed directly by considering its DCT coefficients through (8-10). In (8) and (9) terms on second sigma's argument are independent of input signal, so they only need to calculate once.

### III. PROPOSED METHOD: SML BASED IMAGE FUSION IN DCT DOMAIN

In multi-focus images, the focused area have clear details and are more informative. SML computes higher order changes in both x and y direction. Therefore, clear details of each region is corresponding to higher SML value. So we can use SML as activity level measure of each blocks of the source image. In previous section we show that SML can be calculated from DCT coefficients of block. In this section we use this property to drive our image fusion method based on SML.

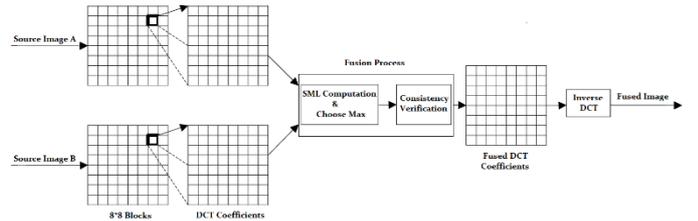

Figure 1. Schematic diagram of fusing images coded in JPEG format

Suppose that A and B are the outputs of two cameras which have been compressed in JPEG standard. Then we have access to DCT coefficients of each $8 \times 8$ blocks of A and B. Fig. 1 shows the schematic diagram of the proposed multi-focus image fusion method. For simplicity, we only consider two source images A and B, but the method can be extended for more than two source images.

The fusion process consists of the following steps:

- Divide source images into blocks of size 8×8. Denote the block pair at location $(i,j)$ by $A_{i,j}$ and $B_{i,j}$ respectively.
- Compute SML of each block by (8-10), and denote the results of $A_{i,j}$ and $B_{i,j}$ by $SMLA_{i,j}$ and $SMLB_{i,j}$, respectively.
- Compare the SML values of two corresponding blocks to decide which should be used to construct the fused image. Create a decision map M to record the feature comparison results according to a selection rule:

$$M_{i,j} = \begin{cases} 1 & SMLA_{i,j} > SMLB_{i,j} + T \\ -1 & SMLA_{i,j} < SMLB_{i,j} - T \\ 0 & otherwise \end{cases} \quad (11)$$

Where, T is a user-defined threshold.

- Apply a consistency verification process to improve quality of the output image. Use a 3×3 majority filter [8] to obtain a refined decision map S:

$$S_{i,j} = \sum_{x=i-1}^{i+1} \sum_{y=j-1}^{j+1} M_{x,y} \quad (12)$$

Then obtain the DCT representation of the fused image F based on R as:

$$F_{i,j} = \begin{cases} A_{i,j} & S_{i,j} > 0 \\ B_{i,j} & S_{i,j} < 0 \\ (A_{i,j} + B_{i,j})/2 & S_{i,j} = 0 \end{cases} \quad (13)$$

- Use inverse DCT to obtain the fused image.

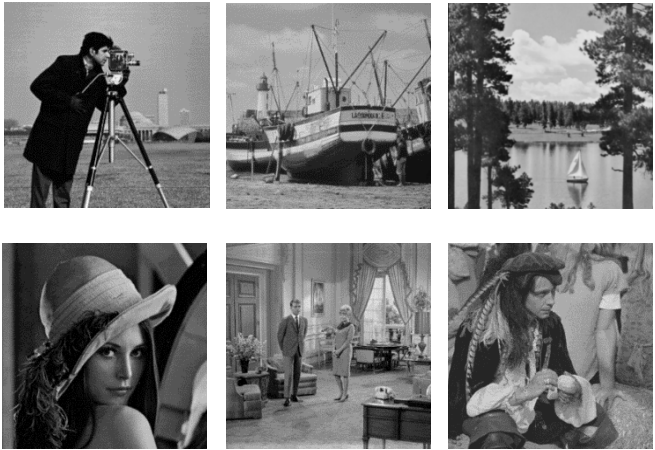

Figure 2. Standard test images used for simulations.

IV. EXPERIMENTAL RESULTS AND ANALYSIS

In this section experimental results of proposed method are given and evaluated by 5 other techniques. Three of them are DCT based techniques and two remaining are DWT and SIDWT based techniques described briefly in section 1.

TABLE I. AVERAGE SSIM VALUES OF ALGORITHMS

| Average SSIM Values | |
|---|---|
| *Algorithm* | *Average SSIM Value* |
| DWT | 0.98389 |
| SIDWT | 0.98112 |
| DCT+Variance [12] | 0.9765 |
| DCT+Variance +CV[12] | 0.9987 |
| DCT+AC_Max[13] | 0.996 |
| DCT+SML(proposed) | **1.0000** |

*A. Performance Measure*

Although there have been many attempts, no universally accepted criterion has emerged yet for objectively evaluating image fusion performance. This problem partially lies in the difficultly of defining an ideal fused image. In this paper we use three performance metrics. One of them are used for referenced images. In this case Structural Similarity (SSIM) [19] are used as quality metric. Here the out-of-focus images are artificially created by low pass filtering of some areas in ideal image. On the other hand Petrovic metric ($Q^{AB/F}$) [20] and Mutual Information (MI) [12] are used to evaluate proposed algorithm on non-referenced multi-focus images. Here non-referenced databases of source images are created by our self.

*B. Experimental Results*

The simulations of the fusion methods have been conducted with an Intel i5 2410 processor with 4 GB RAM. For the wavelet based methods, the DWT with DBSS (2,2) and the SIDWT for Haar basis with three levels of decomposition are applied.

The proposed fusion algorithm is applied on set of non-referenced and set of referenced images and the results are evaluated. The first experiment is conducted using an extensive set of artificially generated images with different focus levels. Eighteen couples of artificial source image have been created by blurring six standard original images shown in Fig. 2, with three disks of different radiuses 5, 7 and 9 pixels. The image are balanced in the sense that the blurring occurs in the both left and right halves of the image. The average SSIM values of 18 experiments are given in Table 1. From the average SSIM values, it can be clearly seen the superior performance of proposed algorithm. In all cases, proposed algorithm results in ideal fused image with SSIM value 1.

The subjective test of the resultant images approves objective results. Due to the lack of space the resultant image of only one image set is shown in Fig. 3. Based on this results, DWT results in a severe blocking effect on fused image. SIDWT method spreads blurring in all areas of image and changes contrast. All other DCT based methods except proposed method, suffer from blocking artifacts especially on the boundaries of focused and defocused regions. Proposed method results in ideal fused image owing to the superior performance of SML criterion.

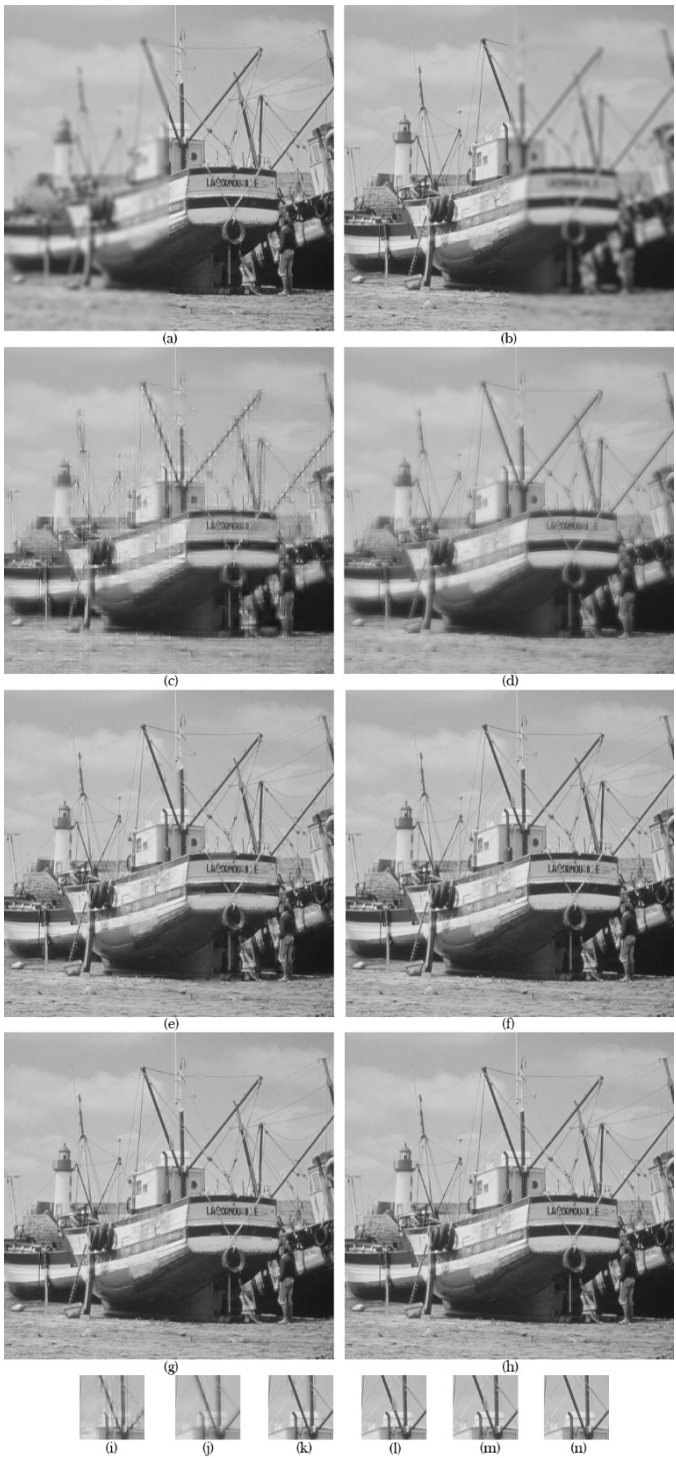

Figure 3. Source images "boats" and the fusion results. (a) the first focus image with focus on the right, (b) the second focus image with focus on the left, (c) DWT result, (d) SIDWT result, (e) DCT+Variance result, (f) DCT+Variance+CV result, (g) DCT+AC_Max Result (h) proposed algorithm result DCT+SML. (i), (j), (k), (l), (m) and (n) are magnified versions of (c), (d), (e), (f), (g) and (h) respectively.

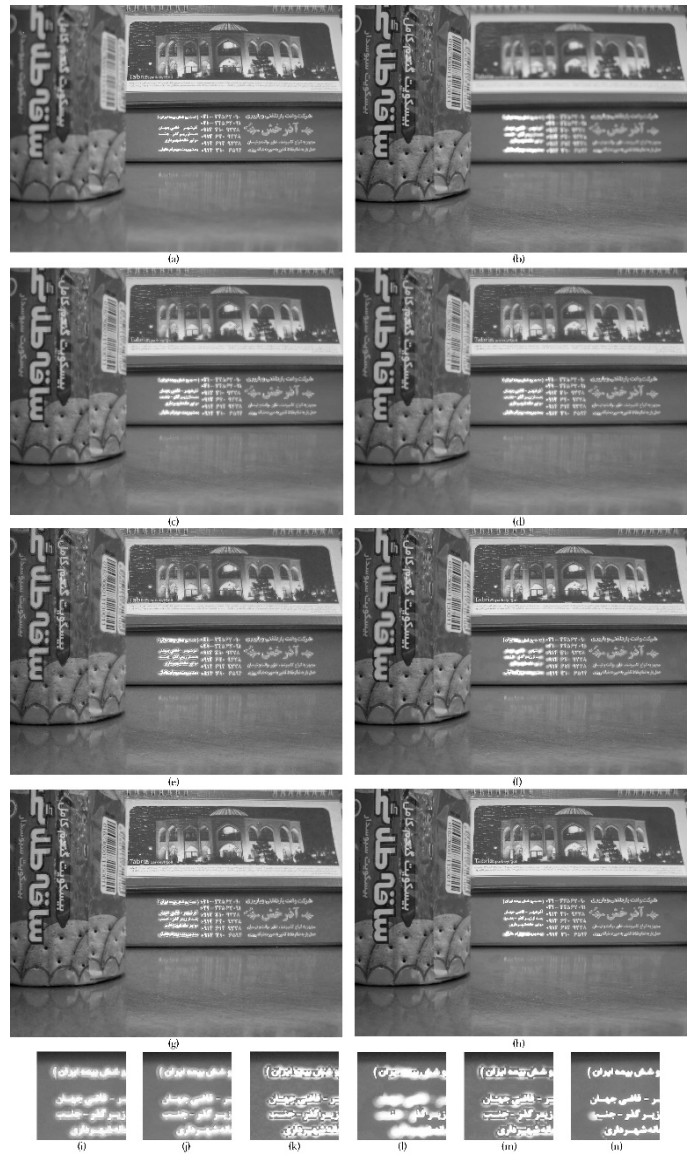

Figure 4. Captured images "Elgoli" by digital camera and the fusion results: same order as in Figure 3.

In order to have some real application experiments, other experiment was conducted on images of a digital camera. Two images are captured using a camera Olympus SP-500UZ with different manually adjusted focus levels and used in experiments. Similar subjective results are obtained on these experiments as shown in Fig. 4. Eventually corresponding objective results, acquired from Petrovic and MI metrics, are given Table 3. This quantitative evaluations certify the subjective results and show the superior performance of proposed method.

Finally, average run times of each algorithm based on reconstruction time of each 8×8 block of images in the DCT based methods are given in Table 6. From the values of this table, it can be seen that superior performance of proposed algorithm obtained by a little addition of runtime.

TABLE II. OBJECTIVE EVALUATION OF THE IMAGE FUSION ALGORITHMS FOR "ELGOLI" DATABASE IN FIG. 4.

| Objective Evaluation | | |
|---|---|---|
| *Algorithm* | *MI* | $Q^{AB/F}$ |
| DWT | 5.7629 | 0.3452 |
| SIDWT | 6.0129 | 0.3888 |
| DCT+Variance [12] | 7.5806 | 0.6555 |
| DCT+Variance+CV [12] | 7.6966 | 0.6570 |
| DCT+AC_Max [13] | 7.6636 | 0.6652 |
| DCT+SML(proposed) | **7.7081** | **0.6690** |

TABLE III. AVERAGE RUNTIME VALUES OF ALGORITHMS (IN MICROSECONDS PER 8×8 BLOCK)

| Objective Evaluation | |
|---|---|
| *Algorithm* | *Runtime* |
| DCT+Variance [12] | 11.32093 |
| DCT+Variance+CV [12] | 52.32136 |
| DCT+AC_Max [13] | 33.67209 |
| DCT+SML(proposed) | 69.73902 |

## V. CONCLUSION

In this paper, a new DCT based fusion technique for multi-focus image was proposed. The method is based on the calculation of SML in DCT domain. SML calculation in DCT domain makes image fusion algorithm simple and suitable for real-time applications. Beside simplicity, better quality of fused image is achieved due to superior performance of SML criterion related to other criterions like variance. Numerous simulations results on different data bases show that proposed method outperforms existing DCT based algorithms. In the case of referenced images, fusion results in ideal fused image. On the other hand fusion of non-referenced images results in fused images with both better subjective and objective image quality.